# Robust and Bright Photoluminescence from Colloidal Nanocrystal/Al$_2$O$_3$ Composite Films fabricated by Atomic Layer Deposition

*Milan Palei, Vincenzo Caligiuri, Stefan Kudera, and Roman Krahne**

Nanochemistry Department, Istituto Italiano di Tecnologia, Genova, Italy

Dipartimento di Chimica e Chimica Industriale, Universita di Genova, Genova, Italy.

AUTHOR INFORMATION

**Corresponding Author**

\* Roman.Krahne@iit.it

Abstract:

Colloidal nanocrystals are a promising fluorescent class of materials, whose spontaneous emission features can be tuned over a broad spectral range via their composition, geometry and size. However, towards embedding nanocrystals films in elaborated device geometries, one significant drawback is the sensitivity of their emission properties on further fabrication processes like lithography, metal or oxide deposition *etc*.. In this work, we demonstrate how bright emitting and robust thin films can be obtained by combining nanocrystal deposition from solutions via spin-coating with subsequent atomic layer deposition of alumina. For the resulting composite films, the layer thickness can be controlled on the nanoscale, and their refractive index can be finely tuned by the amount of deposited alumina. Ellipsometry is used to measure the real and imaginary part of the dielectric permittivity, which gives direct access to the wavelength dependent refractive index absorbance of the film. Detailed analysis of the photophysics of thin films of core-shell nanocrystal with different shape and different shell thickness allow to correlate the behavior of the photoluminescence and of the decay life time to the changes in the non-radiative rate that are induced by the alumina deposition. We show that the photoemission properties of such composite films are stable in wavelength and intensity over several months, and that the photoluminescence completely recovers from heating processes up to 240 °C. The latter is particularly interesting, since it demonstrates robustness to the typical heat treatment that is needed in several process steps like resist-based lithography and deposition by thermal or electron beam evaporation of metals or oxides.

**Introduction**

Colloidal nanocrystals have recently drawn significant interest for applications in light emitting devices due to their size and shape dependent optical and electronics properties, like narrow emission band and ultra-high photoluminescence quantum yield (PLQY).[1-4] In particular, core-shell structures are an efficient structure to reduce the influence of surface defects and allow for more bright and stable emission.[5-10] However, the stability of nanocrystals in thin films is significantly reduced, and their degradation caused by device fabrication processing and over time still hinders their use in effective optoelectronic devices.[11,12] In order to overcome this limitation, several methods have been studied, among which embedding the colloidal nanocrystals in a polymer matrix[13-15] or coating them with dielectric oxides [16-22] revealed promising. Concerning coating of nanocrystal films with dielectric oxides, Atomic Layer Deposition (ALD) has the advantage that the voids present in the thin films constitute suitable sites for the gaseous precursors acting in ALD.[16] Therefore ALD can efficiently infill these voids, which results in a compact composite film in which the nanocrystals are embedded in a dielectric matrix.[18,19,22] In this work, we deposit alumina via ALD on nanocrystal films consisting of CdSe/ZnS core-shell nanoparticles, called dot-in-dots (DiDs) in the following, and CdSe/CdS dot-in-rods (DiRs). We chose those two materials, because on one hand they are among the most prominent colloidal nanomaterials used for light emission,[23-29] and because their difference in shape and architecture leads to different optical properties.[30,31] For example, DiRs can manifest polarized emission, and their elongated shape results in more porous films obtained by spin-coating. Although the PL intensity of the nanocrystal films is reduced by alumina deposition with just a few ALD cycles, we find that it recovers significantly once the infilling of the film is completed, which typically happens at 20-50 ALD cycles. Furthermore, the refractive index of the composite layer can be tuned to higher values with increasing amount of deposited alumina, until it saturates when the film is fully filled and the overcoating of the nanocrystal layer takes place. Concerning the study of the optical properties of such films, ellipsometry is extremely useful, both for evaluating the refractive index (*via* the real part of the dielectric permittivity) and for measuring the absorbance of the thin nanocrystal films via the imaginary part of the permittivity. The latter is usually problematic to measure in detail in transmission experiments, due to considerable background that originates from light scattering. Finally, composites films, where the nanocrystal layer has been fully overcoated by alumina, are robust to temperature treatment up to 240 °C, and maintain their PL intensity over several months of storage.

**Results and Discussion**

We have prepared thin films of CdSe/ZnS DiD and CdSe/CdS DiR nanocrystals by spincoating them from colloidal solutions on glass substrates. Figures S1-S3 show TEM images of the nanocrystals, and their absorption and emission in solution. After nanocrystal deposition, the samples were inserted in the ALD system and different amounts of alumina were deposited by varying the number of ALD cycles (ranging 5 to 200). For a small number of cycles, we can expect that the alumina fills the voids in the nanocrystal film, as illustrated in Figure 1a, and refer to this process as "infilling" in the following. Once the voids have been completely filled, the ALD of alumina increases the thickness of the layer, which can be related to overcoating of the composite layer with alumina (see Figure 1b). We measured the thickness of the films before and after the alumina deposition with a profilometer, as shown in Figure 1c. We find that up to around 20 cycles the film thickness remains more or less constant, followed by a linear increase. This allows us to distinguish the infilling (flat) from the overcoating (linear increase) range, as illustrated by the shaded area in Figure 1c. The morphology the nanocrystal films was evaluated by scanning electron microscopy (SEM) images (see Figure S4 in the SI), which reveal a higher packing density of the particles in the DiD films with respect to the DiR ones. This higher packing density, which corresponds to a smaller volume of voids, leads to the increase of film thickness at a smaller number of ALD cycles for the DiDs (red) with respect to the DiRs (blue).

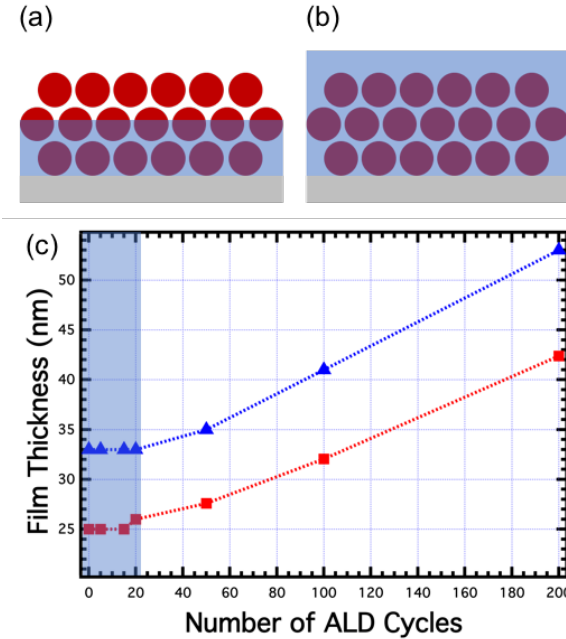

Figure 1. (a,b) Simplified schemes of a partially infilled (a) and an overcoated (b) nanocrystal film on a planar substrate. (c) Film thickness versus number of ALD cycles for DiD (red) and DiR (blue) films. The shaded region indicates the "infilling" regime, in which the alumina is deposited within the nanocrystals in the film, i.e. without significantly increasing the thickness of the layer.

The infilling and overcoating of the nanocrystal layer with alumina can be expected to modify the dielectric properties of the composite film, therefore, we have analyzed the dielectric permittivity by ellipsometry. Figure 2a and Figure 2c show the real part of the dielectric permittivity of the composite nanocrystal/alumina films as a function of the number of ALD cycles, measured at their respective emission wavelength. As expected from filling the voids in the film with a high refractive index material such as alumina, $\varepsilon'$, taken at the maximum of the PL peak, initially strongly increases with the number of ALD cycles (in the infilling regime), and eventually saturates in the overcoating regime. The DiD film manifests a sharp increase in dielectric constant within the first 20 ALD cycles, with a total increase from 3.41 to 3.62. For the DiR sample we observe an equally strong increase of 0.2 in $\varepsilon'$ (from 3.52 to 3.72) in the infilling regime that occurs more gradually, which that can be related to the slower filling of the larger amount of voids present in the DiR nanocrystal film. In this respect, the data obtained by ellipsometry allows us to distinguish the infilling from the overcoating regime more precisely.

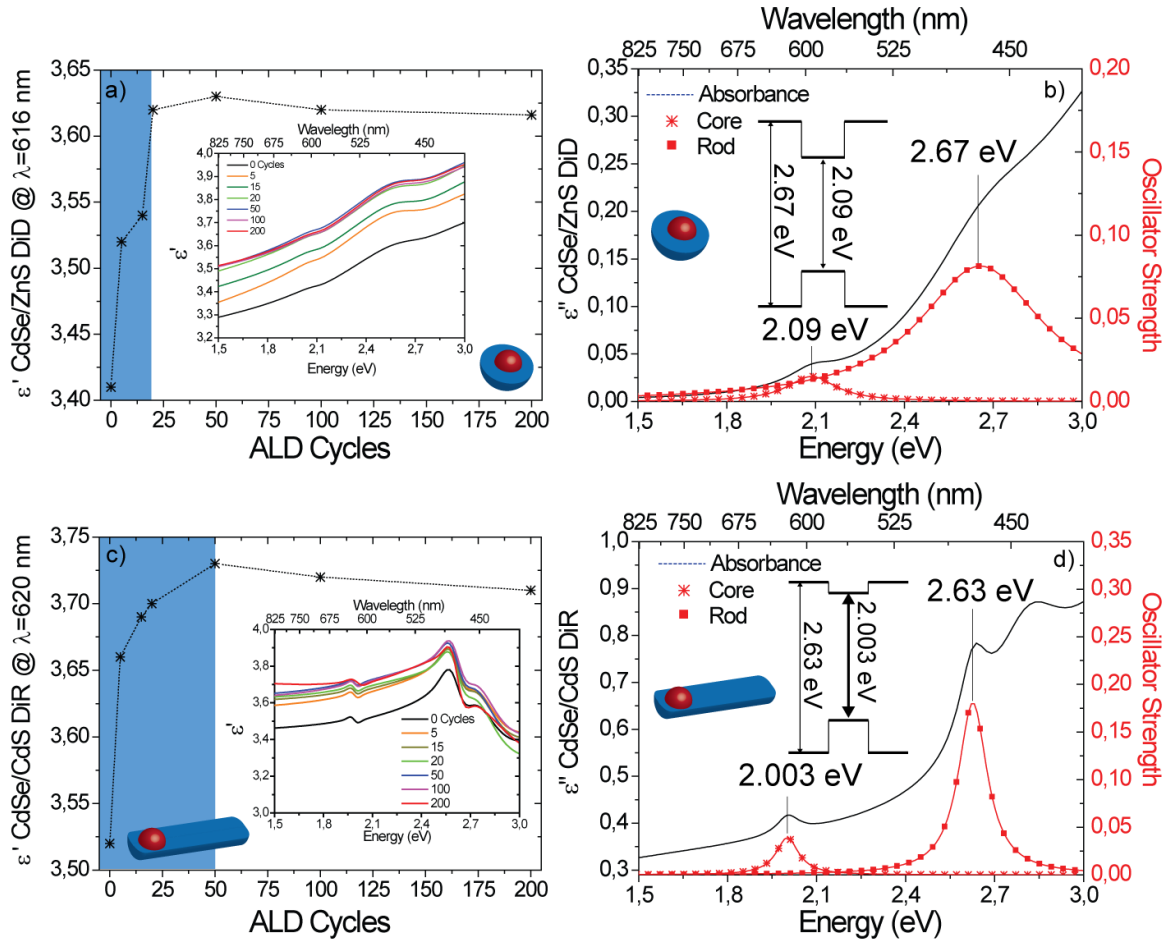

Figure 2. (a) and (c) depict the real part of the dielectric permittivity (ε') acquired at the emission peak wavelength of DiD (a) and DiR (c), as a function of the number of ALD cycles. The shaded area highlights the infilling regime in which the permittivity increases. (b) and (d) show spectroscopic ellipsometry analysis of the imaginary part of the dielectric permittivity (ε'') of the CdSe/ZnS DiD (b) and CdSe/CdS DiR (d) films. The contribution originating from the absorbance of the cores and the shells of the two systems have been fitted by means of lorentzian oscillators.

The absorption spectrum of a thin nancrystal film is often problematic to measure in optical transmission experiments, for example with a spectrophotometer, since the morphology of the film results in considerable scattering. This light scattering typically leads to a pronounced tail in the absorption spectrum that camouflages details in this spectral range, which can make the assignment of the precise onset in absorption difficult. Another option that works extremely well for thin nanocrystal films is to access their absorbance by measuring the imaginary part of complex dielectric permittivity by ellipsometry. Figure S5 shows a comparison of the measured absorption and dielectric permittivity spectra of a composite nanocrystal film. The imaginary part of complex dielectric permittivity provides a direct measure of the absorbance, and allows for a very accurate assignment of the absorbance onset (*i.e.* the optical band gap), and higher energy transitions, as can be seen in Figures 2b and 2d. In particular, the contributions to the total absorbance deriving from the CdSe cores and CdS shells of the nanocrystals are evident in the spectra. These contributions can be quantitatively evaluated by modelling with damped harmonic oscillators, [32-34] which allows fitting of the resonances with the following complex term that contributes to the permittivity:

$$\tilde{\varepsilon}(E) = \frac{\alpha\, E_0}{E_0^2 - E^2 - i\beta E}; \quad (1)$$

Here $E_0$ is the central energy, E is the energy, α is the amplitude at $E_0$, and β is the full-width-at-half-maximum. In terms of the Lorentzian oscillator model, α is the oscillator strength, and β represents the damping of the oscillator. We note that eq. (1) is intrinsically Kramers-Krönig consistent, and that the real imaginary part of the function is equivalent to the description of light absorption in eq. (3) in Jellison et al. [32]. The parameters related to the optical transitions in the CdSe/ZnS DiD and CdSe/CdS DiR nanocrystal films assigned to the core-shell band structure are reported in Table 1.

Table 1. Energy, oscillator strength and broadening of the core and shell transitions of the DiD and DiR nanocrystal films obtained by ellipsometry.

|  | $E_0$ (eV) | α | B |
|---|---|---|---|
| **DiD Core** | 2.09 | 0.015 | 0.199 |
| **DiD Shell** | 2.67 | 0.081 | 0.507 |
| **DiR Core** | 2.003 | 0.039 | 0.093 |
| **DiR Shell** | 2.63 | 0.181 | 0.118 |

Figure 3 shows the emission properties of the different films for CdSe/ZnS DiDs, and CdSe/CdS DiRs with respect to the number of ALD cycles. In all cases, we observe a rapid decrease in PL intensity within the first few cycles of ALD, followed by a recovery, and eventually again a much slower decrease. The maximum in recovery correlates very well with the transition from the infilling to the overcoating regime. For the DiRs, the photoluminescence quantum yield (PLQY) shows a similar behaviour as the PL amplitude, and in the case of the thick-shell DiRs (see Figure S3 for TEM image and optical data recorded from solution phase) it reaches even higher values at 100 ALD cycles as compared to the PLQY of the initial DiR layer. Concerning the DiDs, the PLQY is overall lower and remains more or less constant, within the error margins. The average life time of the emission was evaluated by fitting the decay traces obtained in TCSPC measurements at the peak maximum with a three-exponential function (see the SI for details). For the DiDs and thin-shell DiRs it decreases for few ALD cycles, recovers around 100 ALD cycles and for 200 ALD cycles the average life time is consistently larger than that of the bare nanocrystal layer. For the thick-shell DiR films no decrease, but just an increase of the average life time is observed, and the values of the average life times are much larger than those of the thinner shell nanocrystals. The overall much larger life time of the thick-shell DiRs is in good agreement with those reported in ref. [35]. Next, we calculate the average radiative and non-radiative decay rates using the two equations: [33,36]

$$QY = \frac{\Gamma_{rad}}{\Gamma_{rad} + \Gamma_{non-rad}} \quad (2)$$

$$\tau_{AVG} = \frac{1}{\Gamma_{rad} + \Gamma_{non-rad}} \quad (3)$$

in which $\tau_{AVG}$ is the average lifetime, and $\Gamma_{rad}$ and $\Gamma_{non-rad}$ are the average radiative and non-radiative decay rates, respectively. The average radiative and non-radiative rates are plotted in the lowest panels of Figure 3. We find that the ALD coating mainly affects the non-radiative rate for the DiDs and thin-shell DiRs, while the radiative rate remains more or less constant. For the thick-shell DiRs the radiative and non-radiative rates are only marginally affected by the alumina coating. From this set of data, we can draw the following conclusions. For few cycles of ALD alumina coating the surface passivation of the nanocrystals, initially provided by the organic surface ligands, is deteriorated by the deposition of alumina that occurred at a temperature of 110 °C. Therefore, the non-radiative rate increases for the nanocrystals with a thin shell (DiDs and DiRs in (b)), which results in a decrease PL, PLQY, and average life time. Once the nanocrystal layer is completely filled and slightly overcoated with alumina (at 100 ALD cycles), the nanocrystal surface is again well protected from environmental influences that could quench the emission, and therefore the PL signal has recovered. The magnitude of the recovery depends on the quality of the surface passivation in the composite film, and on the sensitivity of the nanocrystal emission to surface effects. The DiDs that manifest more tightly confined excitons to the CdSe core region (by a thicker and higher bandgap shell) as compared to the thin-shell DiRs show an almost complete recovery of their PL (up to 96%), while the thin-shell DiRs recover only 70 % of their original PL. Also, the recovery of the PL in the DiD films occurs at lower number of ALD cycles (50 for DiDs, while 100 for DiRs), which directly relates to the faster infilling due to the much smaller void volume. The increase in PLQY and the much smaller impact of the alumina coating on the non-radiative rate in the case of thick-shell DiRs can be related to their increased insensitivity to surface modifications[24] and their long average PL decay times[35]. We also coated a monolayer of DiDs that was fabricated by Langmuir Blodgett deposition with alumina by ALD in order to evaluate the influence of the NC film thickness on the emission properties. The results are depicted in Figure S7 and overall show a similar behaviour as the NC films described above, although the recovery in PL intensity reaches only up to 80 % and occurs slightly earlier (at 40 cycles). Furthermore, overcoating of thin-shell DiR films, on which the native trioctylphosphine oxide (TOPO) or octadecylphosphonic acid (ODPA) ligands have been exchanged with aminoethanethiol (AET),[28] manifest a strong reduction in PLQY upon ligand exchange, and further a continuous reduction in PL intensity with the number of ALD cycles. This behaviour underlines the importance of the ligand passivation, and indicates that ligands on the NC surface are not removed or carbonized by the ALD that was performed at a temperature of 110 °C. Indeed, film treatment at 110 °C under vacuum should lead to the removal of excess ligands in the film, but not remove the ligands that passivate the NC surface.[37] This is further corroborated by Fourier-Transform infrared (FTIR) spectra shown in Figure S8 which do not manifest significant differences with and without alumina deposition.

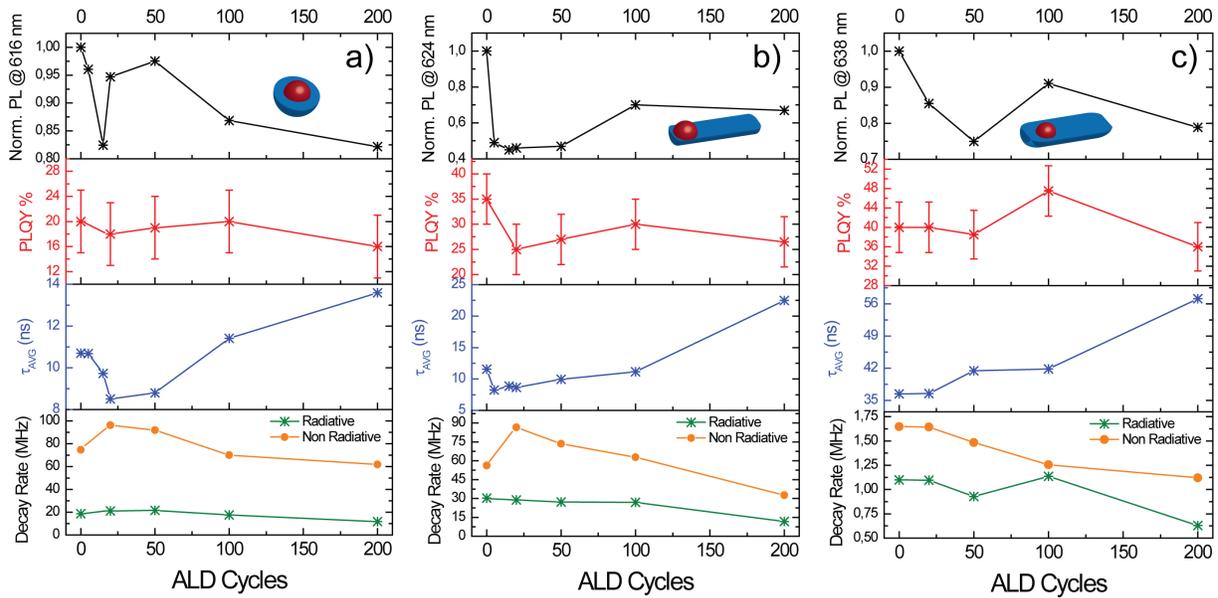

Figure 3. Photo-physical properties of DiD (a), thin-shell DiR (b), and thick-shell DiR (c) films coated with alumina via different number ALD cycles. The PL amplitude at the peak maximum, the PLQY, average life time, and the radiative and non-radiative decay rates are reported. PL spectra are displayed in Figure S6 in the SI.

The overcoating of the nanocrystal films with alumina should lead to an improved stability of the optical properties towards harsh environmental conditions. Figure 4 shows the PL intensity recorded versus temperature in a heating and cooling cycle from 300 K to 520 K, normalized to its initial value at room temperature (300 K). In all cases the PL intensity strongly decreases with increasing temperature, following an Arrhenius behavior. However, for the non-coated films we observe also a significant reduction of the PL at RT after a completed temperature cycle that can be related to heat-induced degradation of the nanocrystals in the film, or of their surface passivation. On the contrary, the alumina overcoated nanocrystal films are robust to this temperature treatment and maintain over 80 % of the PL for the DiDs, and 100% for the DiR after the cycle at RT. We tentatively attribute the slight decrease in PL observed for the alumina coated

DiD films to temperature induced changes at their surface, since the octadecylamine ligands used for these commercially available CdSe/ZnS might suffer from degradation at this temperature.

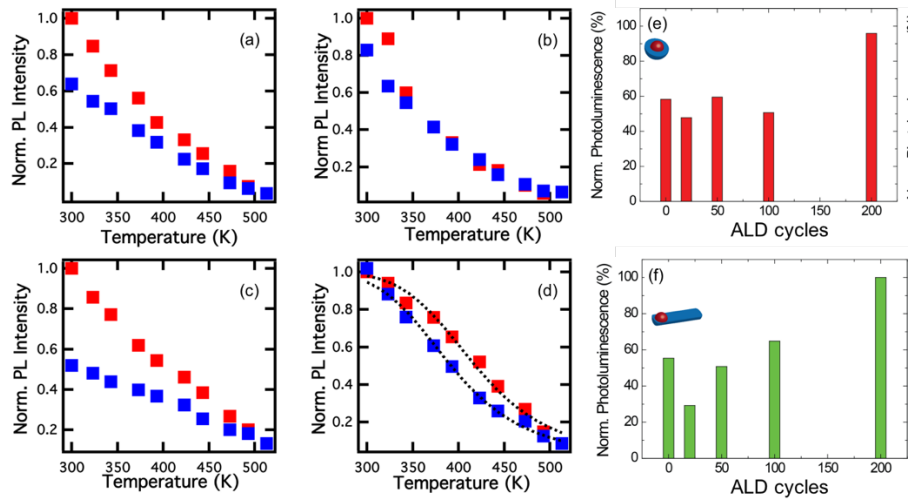

Figure 4. PL intensity versus temperature measured under ambient conditions for (a) non-coated DiD films, (b) DiD films coated by 200 ALD cycles of alumina, (c) non-coated DiR films and (d) DiR films coated by 200 ALD cycles of alumina. (e,f) PL intensity recorded after 3 months of shelf time for nanocrystal films of DiDs (e) and DiRs (f) that were coated with different number of ALD cycles. The PL is normalized to the one measured directly after film fabrication. The PL spectra related to (a-d) are shown in the SI in Figures S9 and S10.

The robustness of the alumina overcoated DiR film to the heating up to 520K allows for fitting of the data with the Arrhenius formula:[38-41, 42]

$$I(T) = \frac{1}{1+C\,exp\left(-\frac{E_A}{k_B T}\right)} ; \quad (4)$$

Where $I(T)$ is the normalized PL intensity, $C$ is a constant taking the non-radiative recombination into account, $E_A$ is the activation energy related to the non-radiative recombination process, and $k_B$ is the Boltzmann constant. The fitting in Figure 4d yields an activation energy of around 0.3 eV, which is in reasonable agreement with the trap depths estimated for CdSe/CdS core-shell structures in ref. [39]. The deposition of approximately 20 nm of alumina by 200 ALD cycles effectively protects the NC film from degradation due to oxidation at elevated temperatures, which indicates an effective sealing from the gases present in ambient atmosphere.

The ALD coating process with alumina is also able to protect the nanocrystal surface from degradation under ambient conditions for extended periods of time. Figures 4 (e,f) show the PL peak intensity, normalized to its value measured directly after film preparation, after 3 months of sample storage for different number of ALD cycles, i.e. for different amounts of alumina deposition that result in infilling and overcoating of the nanocrystal layer. While partial filling of the nanocrystal layer does not increase the shelf stability, we clearly observe that overcoating the nanocrystal layer with 200 ALD cycles of alumina leads to stable emission intensity even after three months of sample storage. On the other hand, ALD infilling and overcoating of the NC layers results in a reduction of the conductance of the film in vertical configuration. Figure S11 in the SI shows the current versus bias voltage of a DiD layer for different number of ALD cycles of alumina, where the conductance decreases exponentially with increasing ALD deposition. This is detrimental for devices based on electroluminescence, but not for applications in down-conversion.

**Conclusion**

In conclusion, we reported the fabrication of composite colloidal nanocrystal/alumina films in which the infilling and overcoating regimes of the nanocrystal layer can be controlled precisely identified by ellipsometry. In the infilling regime, the dielectric constant of the composite film can be tuned by the amount of deposited alumina. PL intensity and decay dynamics recover close to their original values for alumina overcoated films. The alumina overcoated nanocrystal films manifest robust and stable PL emission, both over time in with respect to heating of the film. This improvement of the film stability enables post-processing of the films without deteriorating their emission, for example by lithography techniques and metal deposition. Together with the opportunity to tailor the optical properties like refractive index, these properties are highly advantageous for controlling, for example, the coupling to plasmonic structures that potentially can be fabricated on top of the overcoated films, and which can lead to increased performance in LEDs or in energy harvesting. Furthermore, robust layered structures are promising for applications in metamaterial based nano-lasers, nano-lenses and sensors where highly emissive and high-refractive index gain materials are required.

*Materials and Methods*

Nanocrystal materials: CdSe/ZnS DiDs stabilized with octadecylamine ligands were purchased from Sigma Aldrich (product number 790192) and dispersed in toluene with a concentration of of 5mg/mL. For the thin-shell CdSe/CdS DiRs, first the CdSe cores with diameter of 4.3 nm were synthesized and purified according to the procedure described by Carbone et al..[9] For the shell growth, 90 mg of cadmium oxide was mixed together with 480 mg of ODPA, 3 mg TOPO and 60 mg HPA in a flask, and degassed at 130 $^{0}$C for an hour. Then 100 μM of the CdSe cores dispersed in S:TOP(90 mg:1.5 gm) were swiftly injected at 365 $^{0}$C into the flask under nitrogen atmosphere. After 8 minutes, the growth was terminated by cooling down the reaction. The samples were purified by precipitation with methanol, followed by centrifugation and resuspension in toluene. TEM images and optical properties of these DiRs are shown in Fig. S2.

The thick-shell DiRs were synthesized from CdSe cores with 4.2 nm diameter according to the procedure described in ref. [28]. These DiRs had a diameter of 9 nm and a length of 17 nm (see Figure S3).

Film preparation: The DIR and DID films were spin-coated on glass substrates at 3000 rpm for 1 minute, which resulted in a thickness of 33±3 nm and 25±2 nm for DIR and DID film respectively. The film thickness was measured with a Dektak profilometer.

Monolayer film preparation:

Langmuir Blodgett troughs made from teflon with a volume capacity of 15 mL were filled with 8 mL of diethylene glycol. 20 $\mu$L of DiD NCs solution in hexane (1 mg/mL) were drop-cast on the top of diethylene glycol and the wells were covered for 10 minutes to ensure slow evaporation of hexane. Then the cover was removed, and the residual hexane was left to evaporate. After this, a yellow-colored floating film was formed over diethylene glycol. The film was fished from the solution with a glass substrate of size 5×5 mm$^2$, and dried on a hot plate at 80 °C to remove any remaining glycol.

ALD of alumina: Atomic layer deposition was carried out in a Flexal ALD system from Oxford Instruments, using a thermal process with a stage temperature of 110 °C, resulting in an alumina deposition rate of 0.09 nm/cycle. Tri-methylaluminate (TMA) and $H_2O$ were used as precursors, and we performed a pre-heating step for 300 s before stating the ALD cycles. Each cycle consisted of a $H_2O$/purge/TMA/purge with a pulse durations of 0.075/6/0.033/2 seconds, respectively.

Ellipsometry: Spectroscopic Ellipsometry has been performed by with a VVASE ellipsometer by Woollam in the range from 300 nm to 900 nm. Spectroscopic analysis, reflectance and transmittance spectra have been acquired at three different angles (50, 60, and 70°). Additionally, in order to refine the fit, normal incidence transmittance has been collected as well and fitted in the model. All measurements have been conducted with a resolution of 3 nm, and have been normalized to the intensity of the Xe lamp.

Optical spectroscopy: PL on nanocrystals in solution and in films were carried out with an Edinburgh Instruments fluorescence spectrometer (FLS920), which included a Xenon lamp with monochromator for steady-state PL excitation excitation. The PL spectra recorded from films were obtained with an excitation wavelength of 400 nm by keeping excitation and emission slit width fixed at 2 nm and 1 nm, respectively. The spectra were recorded with 2 nm resolution and 0.5 second dwell time. Time resolved PL decay was studied with a time-correlated single-photon-counting unit coupled to a pulsed diode laser. The PL decay traces were recorded by exciting the sample at 405 nm with a 50 ps pulse at a repetition rate of 0.05 – 1 MHz. The signal was collected at PL peak wavelength with a bandwidth of 10 nm. The PLQY of the films has been measured with a calibrated integrating sphere and an Edinburgh FLS900 fluorescence spectrometer. All the measurements were carried out on spin-coated NCs film on glass substrate with dimensions of 1.6×1.3 $cm^2$ at an excitation wavelength of 400 nm. The optical density of the films was kept around 0.1 at the excitation wavelength. The signal was collected with an excitation and emission bandwidth of 10 and 0.40 nm, respectively.

Heat treatment: The substrates were placed on a hot plate under ambient conditions and the PL was collected, under excitation at 350 nm with diode laser, with a lens and optical fiber coupled to an Ocean Optics (HR4000) spectrometer.

ACKNOWLEDGMENT

The research leading to these results has received funding from Horizon 2020 under the Marie Skłodowska-Curie Grant Agreement COMPASS No. 691185, and from the MSCA project 659144 'SCEL-TA'. We thank Dr. Prachi Rastogi for providing the DiR nanocrystal material.

**Supporting Information Available**. (1) TEM images, absorption and emission spectra of the nanocrystals; (2) SEM images and absorption spectra of the composite films; (3) PL spectra of the films for different number of ALD cycles; (4) FTIR spectra; (5) Temperature-dependent PL spectra; (6) Vertical conductivity measurements;

SYNOPSIS TOC

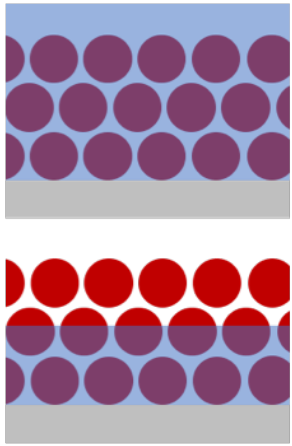
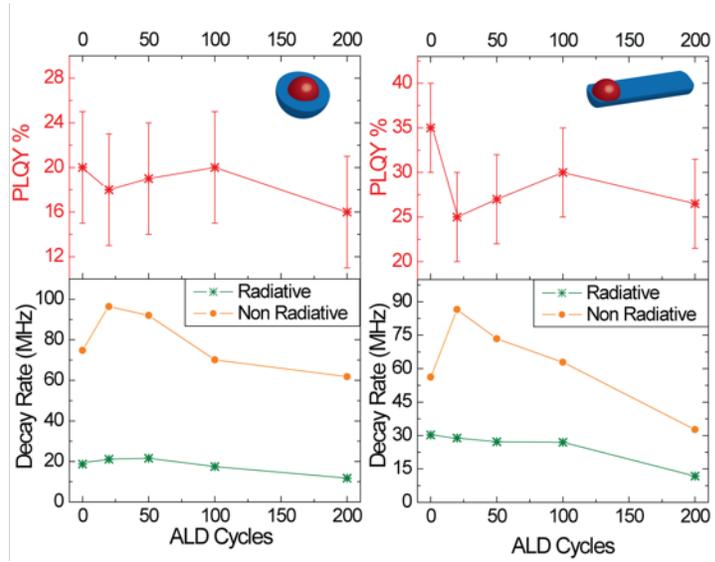